\documentstyle[12pt,fleqn,epsf]{article}
\def\beq{\begin{equation}}
\def\eeq{\end{equation}}
\def\bea{\begin{eqnarray}}
\def\eea{\end{eqnarray}}
\def\etal{\hbox{\it et al.}}
\def\np#1#2#3{    {\it Nucl. Phys. }{\bf #1} (19#2) #3}
\def\pl#1#2#3{    {\it Phys. Lett. }{\bf #1} (19#2) #3}
\def\pr#1#2#3{    {\it Phys. Rev. }{\bf #1} (19#2) #3}

\def\prl#1#2#3{    {\it Phys. Rev. Lett. }{\bf #1} (19#2) #3}

\def\zp#1#2#3{    {\it Zeit. f\"ur Physik }{\bf #1} (19#2) #3}

\def\prepnuma{\phantom{X}}
\def\prepnumb{\phantom{X}}
\def\mxfigura#1#2#3#4{
  \begin{figure}[hbtp]
    \begin{center}
      \epsfxsize=#1
      \leavevmode
      \epsffile{#2}
     \end{center}
    \caption{#3}
    \label{#4}
  \end{figure} }
\def\titol{The Weak-Magnetic Moment of Heavy Quarks}
\def\autora { J. Bernab{\'e}u, J. Vidal}
\def\adressaa{Departament de F{\'{\i }}sica Te{\`o}rica, 
Universitat de Val{\`e}ncia \\ and IFIC, Centre Mixt Univ. Valencia-CSIC\\
E-46100 Burjassot (Val\`encia), Spain}
\def\autorb{G.A. Gonz{\'a}lez-Sprinberg}
\def\adressab{Instituto de F\'{\i}sica, Facultad de Ciencias,\\
Universidad de la Rep\'ublica, CP 10773\\ 11200 Montevideo, Uruguay}
\def\resum{
With initial and final particles on-shell, the anomalous weak-magnetic 
dipole moments of  $b$ and $c$ quarks are electroweak 
gauge invariant quantities of the effective couplings
$Zb\bar{b}$ and $Zc\bar{c}$, respectively, and good candidates
to test the Standard Model and/or  new physics. 
Here we present a complete computation of these quantities 
within the Standard Model. We show that decoupling properties
with respect to heavy particles do take place in the weak magnetic moment.
The obtained  values, 
$a_b (M_Z^2)= (2.98-1.56i)\times 10^{-4}$ and 
$a_c (M_Z^2)= (-2.80+1.09i)\times 10^{-5}$
are dominated by  one-gluon exchange diagrams. The electroweak 
corrections are less than $1$\% of  the total magnitude.
}
\def\firstpage{\begin{titlepage}
\baselineskip 0.50cm \null
\vspace*{-1cm}
 \hfill \prepnuma\\
\null \hfill \prepnumb\\
\baselineskip 0.75cm 
\vskip 1.5cm
\begin{center}
{\Large \bf \titol}
\vskip 1.5cm
\baselineskip 0.6cm
{\bf \autora}

{\it \adressaa}
\vskip .5cm
and
\vskip .5cm
{\bf \autorb}

{\it \adressab}
\vskip 1cm
{\sc abstract}
\end{center}
\baselineskip 0.60cm
\begin{quotation} \resum
 \end{quotation}
 
\end{titlepage}\baselineskip 0.75cm}
 
\begin{document}
\firstpage
\setcounter{page}{2}
%
%

\section{Introduction}
The neutral current sector of the Standard Model (SM) has been subjected to
a detailed precision scrutiny in the past few years \cite{particle}. This
has led to establish definite quantum electroweak corrections to an
impressive list of physical observables which see their tree-level values
modified at the percent level. The agreement between the experiment and the
theory proves the correctness of the SM and the machinery of renormalization
in the quantum field theory. Although the issue is not still 
close, it seems \cite{aleph} that even the
$Z$-vertex to heavy quarks, which contains non-decoupling effects \cite{bps}, 
is in agreement with the SM. An alternative to
this methodology consists in isolating {\it new} observables in the quantum
theory which were absent in the tree-level Lagrangian. In this paper we
study the {\it anomalous} weak-magnetic moment (AWMM) of heavy quarks.

The anomalous weak-magnetic moment of fermions carries important
information about their interactions with other particles. It may be seen
as the coefficient of a
chirality-flipping term in the effective Lagrangian of the $Z$ coupled to
fermions. Therefore, at $q^2\neq 0$, it is expected to be proportional to the
mass of the fermion, and only heavy fermions (leptons or quarks)
are good candidates to have a measurable anomalous weak-magnetic moment.
The already mentioned chirality properties indicate that some
insight into the mechanism of mass generation may be obtained from it.
These properties have also been  considered in the context of extended models
\cite{concha}.
In previous work \cite{nos} we have studied  the case of the $tau$ and
shown that it is possible to construct polarization observables
sensitives to the  AWMM. In this paper we focus on quarks, in
particular on the AWMM of the $b$- and $c$-quarks. In Ref. \cite{kim}  
different strategies to 
 detect polarization effects for the $b$-quark are suggested and discussed, 
 so that the observables may become feasible in the future. 

\section{Anomalous Weak-Magnetic Moment }

As the  AWMM is proportional to the mass of the particle, in principle, only
heavy fermions might have a sizeable value for it. The heaviest quark, the 
$top$-quark \cite{top}, would seem to be the perfect
candidate.
The problem arises there in the electroweak gauge invariant properties of
the
defined form factor. As it is
already well known \cite{nos,fuj}
only the on-shell definition of the AWMM is  electroweak gauge invariant and
free of uncertainties.  Nevertheless, recently some
procedures to move off-shell the gauge invariant form factors have been
proposed \cite{papa},  but their invariant properties and physical
significance are still under discussion \cite{latira}. In this paper
we concentrate ourselves in the study of the
AWMM for the heavy quarks produced from on-shell $Z$'s, i.e.
bottom $b$ and charm $c$ quarks. This is of order $\alpha_s$-strong or 
$\alpha$-electroweak radiative
correction  to the $Zq\bar{q}$ vertex.

Using Lorentz covariance, 
the matrix element of the $i$-quark vector neutral current can be 
written in the form:
\beq
\bar{u}_i(p) \,V^{\mu}(p,\bar{p})\, v_i(\bar{p})
=e\  \bar{u}_i(p) \left[ \frac{v_i(q^2) }{2 s_w c_w}\gamma^\mu+i 
\frac{a^w_i(q^2)}{2 m_i}
\sigma^{\mu\eta} q_\eta\right] v_i(\bar{p})
\label{mu}\eeq
where $q^2=(p+\bar{p})^2$ is 
the 4-momentum squared in the center of mass frame,  $e$
is the proton charge and  $s_w$, $c_w$
are  the weak mixing  angle sine and cosine,
respectively. The first term $v_i(q^2)$ is the Dirac
vertex (or charge radius of the fermion $i$) form factor
and it is present at tree level with a value
$v_i(q^2)=T_{i\, 3}-2Q_i\, s_w^2$, whereas the second form factor,
$a_i^w(q^2)$, is the AWMM and it appears due to quantum corrections. 
As already mentioned, at $q^2=M_Z^2$, 
it is a linearly independent and gauge invariant form factor of 
the Lorentz covariant matrix element, contributing
to the physical $Z\longrightarrow q\, \bar{q}$ decay amplitude.

\mxfigura{8cm}{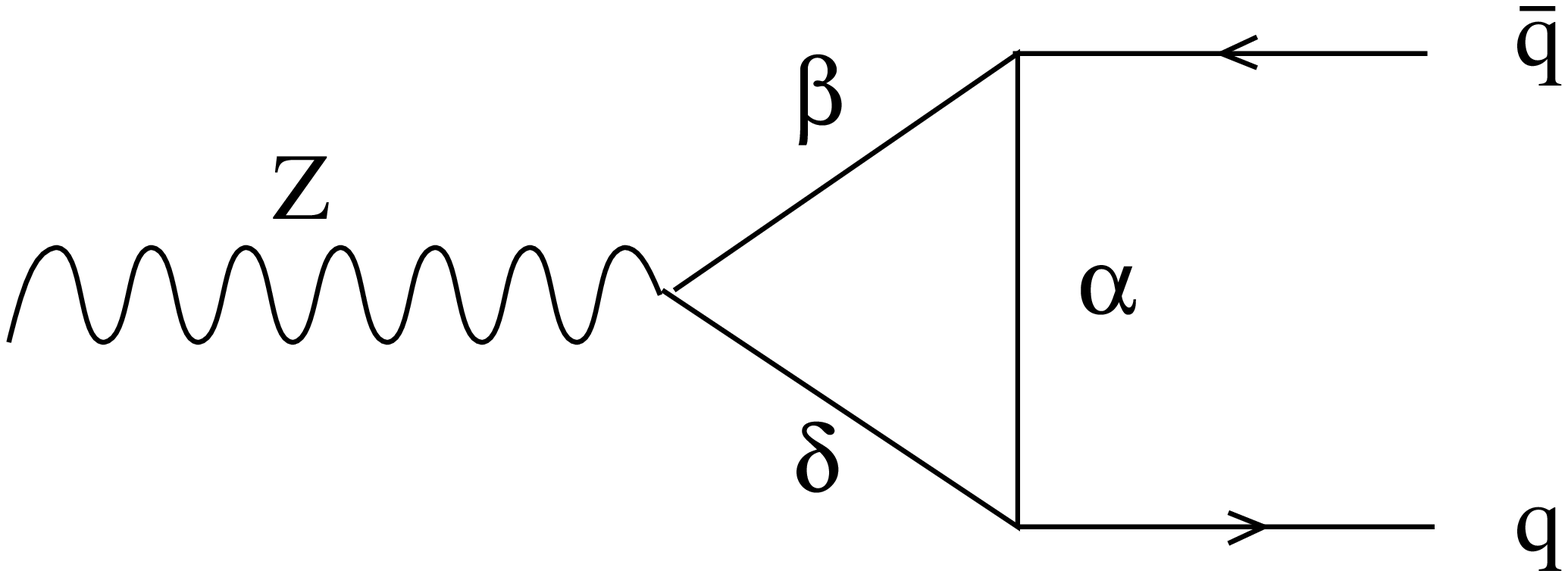}{Contributing Feynman diagrams to the anomalous 
weak magnetic moment.}{fig1}

In the t'Hooft-Feynman gauge, 
there are 14 diagrams that contribute to $a_i^w$. In 
Fig. \ref{fig1} we show the generic one-loop diagram contribution. From 
now on
we denote by $q_i$  ($q_I$)  the internal quark in the loop with the
same (different) charge as the external quark; $\alpha$, $\beta$ and 
$\delta$ are 
the particles circulating in the loop 
as shown in Fig. \ref{fig1};   
$\chi$ and $\Phi$  are the neutral would-be Goldstone boson and 
physical Higgs, and  $\sigma^\pm$ are the charged would-be Goldstone
bosons. Then, all the 
contributions may be written with the compact notation:

\beq
[a_i]^{\alpha\beta\delta}=\frac{\alpha}{4  \pi} \;\;\frac{m_i^2}{M_Z^2}\;\,\sum_{jk} 
c_{jk} {I_{jk}}^{\alpha\beta\delta}(i)
\label{a}\eeq
with $(\alpha\, \beta\, \delta)$ standing
for: $(N\, \bar{q}_i\,  q_i)$, $(C^\pm\, \bar{q}_I\, q_I)$, $(q_I\, C^+\, C^-)$
and $(q_i\, N\, N^\prime)$. $N$ and $N'$ are the neutral particles $\gamma$,
$Z$, $\chi$, $\Phi$ (with $N\neq N'$), and  $C^\pm$ are the charged bosons
$W^\pm$ and $\sigma^\pm$. $c_{jk}$ are coefficients, and 
\begin{equation}
{I_{jk}}^{\alpha\beta\delta}(i) \equiv 
I_{jk}(m_i^2,q^2,m_i^2,m_\alpha^2,m_\delta^2,m_\beta^2)
\end{equation}
are the on-shell ($p^2=m_i^2\,$, 
${\bar{p}\, }^2=m_i^2$ and $q^2=(p+\bar{p})^2=M_Z^2$ ) scalar, vector 
and tensor functions defined, from the one-loop 
3-point functions
\bea
&& I_{00;\ \mu;\ \mu\nu}
\; ({p}^2,(p+\bar{p})^2,{\bar{p}\, }^2,{m_{_A}}^2,
{m_{_B}}^2,{m_{_C}}^2)\,=\nonumber\\[9pt]
&&\hspace*{1cm} \displaystyle{\frac{1}{i \pi^2}}\; \times   \;
\int d^nk 
 \displaystyle{
\frac{\{1\,;\,k_\mu\,;\,k_\mu k_\nu\}}{(k^2-{m_{_A}}^2)
((k-p)^2-{m_{_B}}^2)((k+\bar{p})^2-{m_{_C}}^2)}
} 
\label{point}\eea
as \cite{nos}:  
\bea
I^\mu\,&=&\,({p}-{\bar{p}})^\mu I_{10}+({p}+{\bar{p}})^\mu I_{11} \nonumber\\
I^{\mu\nu}\,&=&\, ({\bar{p}\, }^\mu {\bar{p}\, }^\nu +{p}^\mu {p}^\nu) I_{21}+ 
({\bar{p}\, }^\mu {p}^\nu+ {p}^\mu {\bar{p}\, }^\nu) I_{22}+ 
({\bar{p}\, }^\mu {\bar{p}\, }^\nu -{p}^\mu {p}^\nu) I_{2-1}+g^{\mu\nu} I_{20}
\label{3point}\eea
 
We are only interested in the AWMM  so that, for each diagram,   we have 
to pick up only the $\sigma^{\mu\nu}q_\nu$ coefficient shown in 
Eq.(\ref{mu}).  Though the AWMM receives its leading contribution 
from one loop diagrams (renormalizability excludes 
$\bar{\psi}\sigma_{\mu\nu}\psi\,Z^{\mu\nu}$ terms in the Lagrangian), it is 
finite, and can be extracted from them with no need of 
renormalization. Notice also that only vertex corrections may contribute 
to the AWMM, because the renormalization of the external legs does not 
change the (V-A) Lorentz structure  of the vertex. 

As a test of our calculation we have verified the conservation of the
vector current
\beq
q^\mu \;\bar{u}(p) \,V^\mu(p,\bar{p}) \,v(\bar{p}) \,=\, 0 
\label{cons}\eeq
by explicitly checking that the  coefficient of 
the $q^\mu$ term of the matrix element (\ref{mu}) vanishes. 
This conservation does not occur on each diagram, but it can be confirmed 
by considering cancellations among some of them and, of course, 
in the overall sum.

In the following, we list all  contributions to the AWMM that are 
 written, in a 
self-explanatory notation, as:

\bea
&&\hspace*{-1.cm}\left[a^w_b\right]^{\gamma bb}=-
\left(\frac{\alpha}{4  \pi} \right)
\left(\frac{m_b}{M_Z}\right)^2\frac{4v_b\, Q_b^2 }{s_w c_w} 
\; M_Z^2\left[ I_{10}
+I_{22}-I_{21}\right]^{\gamma bb }\label{gabb}
\eea
\bea
&&\hspace*{-1.cm}\left[a^w_b\right]^{Zbb}=-
\left(\frac{\alpha}{4  \pi} \right)
\left(\frac{m_b}{M_Z}\right)^2 \frac{v_b}{s_w^3 c_w^3}\, M_Z^2
\; \left[ v_b^2\left(I_{22}- I_{21}+I_{10}\right)\right.+\nonumber\\ 
&&\hspace{6cm} \left. a_b^2\left(3I_{22}-3I_{21}+11I_{10}-4I_{00}\right)
\right]^{Zbb }
\eea
\bea
&&\hspace*{-1.cm}\left[a^w_b\right]^{\chi bb}=-
\left(\frac{\alpha}{4  \pi} \right)
\left(\frac{m_b}{M_Z}\right)^2 
\, \frac{m_b^2}{M_Z^2}\, \frac{v_b}{2s_w^3c_w^3}\, M_Z^2\left[
I_{22}-I_{21}\right]^{\chi bb}
\eea
\bea
&&\hspace*{-1.cm}\left[a^w_b\right]^{\Phi bb}=-
\left(\frac{\alpha}{4  \pi} \right)
\left(\frac{m_b}{M_Z}\right)^2 
\, \frac{m_b^2}{M_Z^2}\, \frac{v_b}{2s_w^3c_w^3}\, M_Z^2\left[
I_{22}-I_{21}+2I_{10}\right]^{\Phi bb}
\eea
\bea
&&\hspace*{-1.cm}\left[a^w_b\right]^{W tt}=-
\left(\frac{\alpha}{4  \pi} \right)
\left(\frac{m_b}{M_Z}\right)^2 
\,  \frac{(v_t+a_t)}{s_w^3c_w}\, \left| V_{tb}\right|^2 \, M_Z^2
\left[I_{22}-I_{21}+3I_{10}-I_{00}\right]^{Wtt}
\eea
\bea
&&\hspace*{-1.cm}\left[a^w_b\right]^{\sigma tt}=-
\left(\frac{\alpha}{4  \pi} \right)
\left(\frac{m_b}{M_Z}\right)^2 \left(\frac{m_t}{M_Z}\right)^2
\,  \frac{1}{2 s_w^3c_w^3}\, \left| V_{tb}\right|^2 \, M_Z^2
\left[v_t\left(I_{22}-I_{21}-I_{10}+\frac{}{}\right.\right.\nonumber\\ 
&&\hspace{1cm}\left.\left.\left(\frac{m_b}{m_t}\right)^2
\left(I_{22}-I_{21}+I_{10}\right)\right)-
a_t\left(1-\left(\frac{m_b}{m_t}\right)^2\right)
\left(I_{22}-I_{21}+I_{10}\right)\right]^{\sigma t t}\label{seis}
\eea
\bea
&&\hspace*{-1.cm}\left[a^w_b\right]^{tWW}=
\left(\frac{\alpha}{4  \pi} \right)
\left(\frac{m_b}{M_Z}\right)^2 
\,  \frac{c_w}{s_w^3}\, \left| V_{tb}\right|^2 \, M_Z^2
\left[I_{10}+2I_{21}-2I_{22}\right]^{tWW}
\eea
\bea
&&\hspace*{-1.cm}\left[a^w_b\right]^{t\sigma\sigma}=-
\left(\frac{\alpha}{4  \pi} \right)
\left(\frac{m_b}{M_Z}\right)^2 \left(\frac{m_t}{M_Z}\right)^2
\,  \frac{1-2c_w^2}{2s_w^3c_w^3}\, \left| V_{tb}\right|^2 \, M_Z^2
\left[I_{00}-I_{22}+I_{21}-3I_{10}+\frac{}{}\right.\nonumber\\
&&\hspace{7cm}\left.\left(\frac{m_b}{m_t}\right)^2
\left(I_{21}-I_{22}-I_{10}\right)\right]^{t\sigma\sigma}\label{siete}
\eea
\bea
&&\hspace*{-1.cm}\left[a^w_b\right]^{b Z\Phi}=-
\left(\frac{\alpha}{4  \pi} \right)
\left(\frac{m_b}{M_Z}\right)^2 
\,  \frac{v_b}{2s_w^3c_w^3}\,  M_Z^2\left[I_{11}-I_{10}\right]^{bZ\Phi}
\label{bzphi}
\eea
\bea
&&\hspace*{-1.cm}\left[a^w_b\right]^{b \Phi Z}=
\left(\frac{\alpha}{4  \pi} \right)
\left(\frac{m_b}{M_Z}\right)^2 
\,  \frac{v_b}{2s_w^3c_w^3}\,  M_Z^2\left[I_{11}+I_{10}\right]^{b\Phi Z}
\label{bphiz}
\eea
\bea
&&\hspace*{-1.cm}\left[a^w_b\right]^{tW\sigma}=-
\left(\frac{\alpha}{4  \pi} \right)
\left(\frac{m_b}{M_Z}\right)^2 
\,  \frac{1}{2s_wc_w}\,  \left| V_{tb}\right|^2 \, M_Z^2
\left[I_{10}-I_{11}\right]^{tW\sigma}
\eea
\bea
&&\hspace*{-1.cm}\left[a^w_b\right]^{t\sigma W}=-
\left(\frac{\alpha}{4  \pi} \right)
\left(\frac{m_b}{M_Z}\right)^2 
\,  \frac{1}{2s_wc_w}\,  \left| V_{tb}\right|^2 \, M_Z^2
\left[I_{10}+I_{11}\right]^{t\sigma W}
\eea
\bea
&&\hspace*{-1.cm}\left[a^w_b\right]^{b\Phi\chi }=
\left[a^w_b\right]^{b\chi\Phi }=0
\eea
with $a_{i,I}$, $V_{Ii}$ being the axial vector $Zq\bar{q}$ coupling,  
and the Kobayashi-Maskawa $q_Iq_i$  mixing matrix element, respectively.

Diagrams with the  Higgs ($\Phi$) and the neutral would-be Goldstone boson 
($\chi$) coupled to the $Z$ only contribute to the axial form factor
and not to the magnetic moment, so that one gets the result of Eq.(19).

The {\it natural} scale of each diagram is
$\left(\frac{m_b}{M_{Z,\Phi}}\right)^2$
but those with an exchange of a Higgs (physical or not) between the
two $b$'s (see $[a_b^w]^{\chi bb}$, $[a_b^w]^{\Phi bb}$) are
suppressed by an extra $\left(\frac{m_b}{M_{Z,\Phi}}\right)^2$
factor coming from the Higgs-$b$-$b$ coupling. For similar reasons, due 
to the high value 
of the top mass \cite{top} one could then think  that
those diagrams with Higgs particles coupled to the $t$-quark
($[a_b^w]^{\sigma tt}$, $[a_b^w]^{t \sigma\sigma}$)
would be the dominant ones. In fact, Eqs. (\ref{seis}) and (\ref{siete})
show the $\left(\frac{m_t}{M_Z}\right)^2$  expected factor, which should make
sizeable the contribution coming from these diagrams. Nevertheless, 
contrary to what happens in the charge radius ($\gamma^\mu$) form factor 
\cite{bps}, where
non-decoupling effects take place,
the behaviour with $m_t$ of the $I_{jk}$ integrals given
in Eqs. (\ref{seis}) and (\ref{siete})
prevents the product $m_t^2\, I_{jk}$ to have a
hard dependence with large $m_t$. An expansion of the scalar functions
$I_{ij}^{t\sigma\sigma}$ and $I_{ij}^{\sigma tt}$ --up to leading order-- in
terms of $1/t\equiv(M_Z/m_t)^2$ gives:
\def\fw{f_w}
\bea
M_Z^2 \, I_{00}^{t\sigma\sigma} &=&\frac{1}{t}\left(\log\frac{c_w^2}{t}+
2\fw-1\right)+{\cal{O}}\left(\frac{1}{t^2}\log\frac{c_w^2}{t}\right)\\[8pt]
M_Z^2\, I_{10}^{t\sigma\sigma} &=& \frac{1}{t} \left( \frac{1}{2} 
\log\frac{c_w^2}{t}+f_w-\frac{1}{4} \right)+
{\cal{O}}\left(\frac{1}{t^2}\log\frac{c_w^2}{t}\right)\\[8pt]
M_Z^2\, I_{21}^{t\sigma\sigma} &=& \frac{1}{t} \left(\frac{1}{3} 
\log\frac{c_w^2}{t}
+\frac{2(1-c_w^2)}{3}\fw-\frac{1}{9}+\frac{2c_w^2}{3} \right) +
{\cal{O}}\left(\frac{1}{t^2}\log\frac{c_w^2}{t}\right)\\[8pt]
M_Z^2\, I_{22}^{t\sigma\sigma} &=& -\frac{1}{t}\left(\frac{1}{6}
\log\frac{c_w^2}{t}+\frac{1+2c_w^2}{3}\fw -\frac{2c_w^2}{3}+\frac{1}{36} 
\right) +{\cal{O}}\left(\frac{1}{t^2}\log\frac{c_w^2}{t}\right)
\eea
\bea
M_Z^2\, I_{00}^{\sigma tt} &=& -\frac{1}{t}+{\cal{O}}\left(
\frac{1}{t^2}\log\frac{c_w^2}{t}\right)\\[8pt]
M_Z^2\, I_{10}^{\sigma tt} &=& -\frac{1}{4t}+{\cal{O}}\left(\frac{1}{t^2}
\log\frac{c_w^2}{t}\right)\\[8pt]
M_Z^2\, I_{21}^{\sigma tt} &=& -\frac{1}{9t}+{\cal{O}}\left(
\frac{1}{t^2}\log\frac{c_w^2}{t}\right)\\[8pt]
M_Z^2\, I_{22}^{\sigma tt} &=& \frac{1}{18t}
+{\cal{O}}\left(\frac{1}{t^2}\log\frac{c_w^2}{t}\right)
\eea
with
$\fw=\sqrt{4c_w^2-1}\, \arctan\left(1/\sqrt{4c_w^2-1}\right)$.
As can be seen from the previous expressions,  only a mild
$(M_Z/m_t)^2 \log\left(m_t/M_Z\right)^2$ dependence is got from
the four diagrams that may give   non-decoupling effects with
the $top$-quark mass. The chirality flipping property of 
the magnetic moment makes the difference with respect to 
the charge radius, where non-decoupling effects are seen. In addition, for
the $[a_b^w]^{tW\sigma}$ and  $[a_b^w]^{t\sigma W}$ amplitudes, the AWMM
selects a product of left and right projectors that
gives no linear contribution on $m_t^2$. 

Adding all these terms, we end up with the following result:
\bea
&&\hspace*{3cm}\left[a^w_b\right]^{(\sigma tt)+(t\sigma\sigma)+ (t\sigma W)+
(tW\sigma)}_{lead. ord. in \, m_t}=\nonumber\\[8pt]
&&=\left(
\frac{\alpha}{4  \pi} \right)
\left(\frac{m_b}{M_Z}\right)^2
\, \frac{1}{2 s_w^3 c_w^3}\, \left| V_{tb}\right|^2 \,
\left[\frac{23}{36}-  \frac{11 c_w^2}{9}
+{\cal{O}}\left(\frac{M_Z^2}{m_t^2}\log\left(\frac{M_Z^2}{m_t^2}\right)
\right)\right]
\eea
\noindent and we  conclude that the non-decoupling of a heavy {\it top}
reduces to a constant term for the AWMM.

The ${I_{jk}}^{\alpha\beta\gamma}$ functions are 
analytically computed in terms of  dilogarithm functions. As a check we confronted the result with a numerical integration in the
$m_b \rightarrow 0$ limit. For $m_t=174$ GeV, $M_Z=91.19$ GeV, $s_w^2=0.232$,
$\alpha=1/127.9$ and $m_b=4.5$ GeV, the following numerical contributions 
for each diagram are found: 

\bea
&&\hspace*{-1.cm}\left[a^w_b\right]^{\gamma bb}\simeq 
\left(\frac{\alpha}{4  \pi} \right)
\left(\frac{m_b}{M_Z}\right)^2
\; (1.10-0.57i)=(1.66-0.87i)\times 10^{-6} 
\eea
\bea
&&\hspace*{-1.cm}\left[a^w_b\right]^{Zbb}\simeq
\left(\frac{\alpha}{4  \pi} \right)
\left(\frac{m_b}{M_Z}\right)^2 
\; (1.6+0.71i)=(2.42+1.07i)\times 10^{-6}
\eea
\bea
&&\hspace*{-1.cm}\left[a^w_b\right]^{\chi bb}\simeq 
\left(\frac{\alpha}{4  \pi} \right)
\left(\frac{m_b}{M_Z}\right)^2 
\; (0.31+4.79i)\times 10^{-4}=(4.69+72.5i)\times 10^{-10}
\eea
\bea
&&\hspace*{-1.cm}\left[a^w_b\right]^{\Phi bb}\simeq 
\left(\frac{\alpha}{4  \pi} \right)
\left(\frac{m_b}{M_Z}\right)^2 
\; (-1.86-5.98i;\; -1.43-1.95i;\; -0.91-0.92i)\times 10^{-3}=\nonumber\\
&&\hspace{2cm}(-2.81-9.07i;\; -2.16-2.96i;\; -1.37-1.40i)
\times 10^{-9}\label{hbb}
\eea
\bea
&&\hspace*{-1.cm}\left[a^w_b\right]^{W tt}\simeq 
\left(\frac{\alpha}{4  \pi} \right)
\left(\frac{m_b}{M_Z}\right)^2 
\; (-0.54)=(-0.81)\times 10^{-6}
\eea
\bea
&&\hspace*{-1.cm}\left[a^w_b\right]^{\sigma tt}\simeq 
\left(\frac{\alpha}{4  \pi} \right)
\left(\frac{m_b}{M_Z}\right)^2 
\; (-0.71)=(-1.07)\times 10^{-6}
\eea
\bea
&&\hspace*{-1.cm}\left[a^w_b\right]^{tWW}\simeq 
\left(\frac{\alpha}{4  \pi} \right)
\left(\frac{m_b}{M_Z}\right)^2 
\; (-2.99)=(-4.53)\times 10^{-6}
\eea
\bea
&&\hspace*{-1.cm}\left[a^w_b\right]^{t\sigma\sigma}\simeq 
\left(\frac{\alpha}{4  \pi} \right)
\left(\frac{m_b}{M_Z}\right)^2 
\; (-0.81)=(-1.22)\times 10^{-6}
\eea
\bea
&&\hspace*{-1.cm}\left[a^w_b\right]^{b Z\Phi}=
\left[a^w_b\right]^{b \Phi Z}\simeq 
\left(\frac{\alpha}{4  \pi} \right)
\left(\frac{m_b}{M_Z}\right)^2 
\; (0.57;\; 0.34;\; 0.22)=\nonumber\\
&&\hspace{7.5cm}
(0.98;\; 0.52;\; 0.33)\times 10^{-6}\label{bzh} 
\eea
\bea
&&\hspace*{-1.cm}\left[a^w_b\right]^{tW\sigma}=
\left[a^w_b\right]^{t\sigma W}\simeq 
\left(\frac{\alpha}{4  \pi} \right)
\left(\frac{m_b}{M_Z}\right)^2 
\; (0.17)=(2.59)\times 10^{-7}
\eea
where the  values between parenthesis in Eqs. (\ref{hbb}) and (\ref{bzh})
correspond to $\frac{M_\Phi}{M_Z}=1,2,3$. The other values agreee with the
result of Ref. \cite{concha} for the SM. We have taken
the Kobayashi-Maskawa matrix being unity ($V_{Ii}=\mbox{diag}(1,1,1)$) 
for numerical results.

Finally, the electroweak  contribution to the $b$-AWMM is
\beq
a_b^w (M_Z^2)= \left[\;- (1.1;\; 2.0;\; 2.4)+0.2i\right] \times 10^{-6}, 
\hspace{1cm} \left[M_\Phi=M_Z,\ 2M_Z,\ 3M_Z\right]
\label{anom}
\eeq

An immediate consequence of these results is that the AWMM 
contribution to the total
electroweak width is very small. This is easily seen just by 
considering that the ratio $\Gamma(a_b^w)/\Gamma^w_{\rm Tree}$ 
is given by the interference with the AWMM amplitude. 
\beq
\frac{\Gamma(a_b^w)}{\Gamma^w_{\rm Tree}}\approx 
6\frac{s_w c_w v_b}{v_b^2+a_b^2}\,  \mbox{Re}(a^w_b)
\eeq
\noindent Then, Eq. (\ref{anom}) shows that only approximately 1 
over $10^{6}$ parts 
of the width is given by the electroweak contribution to the AWMM.

Contrary to what happened for the $tau$ weak magnetic moment, where the lepton
vector neutral coupling was responsible for the suppression of the Higgs mass
dependence, we observe here that the mass of the physical Higgs has a
sizeable effect on the final electroweak magnetic moment (\ref{anom}) for
the $b$-quark. 
For the selected range of $M_\Phi$, it 
changes the real part of the AWMM in more than $100\%$. This is so
because, as can be seen from Eq. (\ref{bzh}), the contribution of 
the $b Z\Phi$ diagrams --Eqs. (\ref{bzphi}) and 
(\ref{bphiz})-- are almost of the same order as the leading ones.
Unfortunately, these effects will not be
observable because, as we will show in the following, the magnetic moment is 
dominated by the QCD contributions.

In addition to the purely electroweak contributions to the AWMM of
the $b$-quark given above, we now consider the QCD contributions to
$a_b$.
To lowest order, there is only one relevant diagram of the type shown in Fig.
\ref{fig1}: the one  with $\alpha$ being now a $gluon$.
The evaluation of that diagram
only differs from the $\gamma b \bar{b}$ diagram (Eq. (\ref{gabb})) 
in the couplings, so that it is straightforward to find the result
\bea
\left[a_b^{QCD}\right]^{gbb}&=&\frac{\alpha_s}{\alpha}\frac{4}{3Q_b^2}
\left[a_b^w\right]^{\gamma b b}=
\left(\frac{\alpha_s}{4  \pi} \right)
\left(\frac{m_b}{M_Z}\right)^2\frac{v_b }{s_w c_w}\, 
\frac{8}{3\beta}\,\left(\log\frac{1-\beta}
{1+\beta}+i\pi\right)\label{qcd}\nonumber\\ 
&=&(2.99-1.56i)\times 10^{-4}\label{glbb}
\eea
with $\beta=\sqrt{1-4(m_b/M_Z)^2}$ and  $\alpha_s=0.117$, which is in
good agreement with the analytical expression found in Ref. 
\cite{arcadi}, when expressed in terms of an AWMM.

The final value we get for the weak magnetic moment of the $b$-quark is then
\beq
a_b(M_Z^2)=a^w_b(M_Z^2)+a_b^{QCD}(M_Z^2)=(2.98-1.56i)\times 10^{-4}
\eeq
\noindent for $M_\Phi=M_Z$.

Eqs. (\ref{anom}) and (\ref{glbb}) show that even though different
values of the Higgs mass modify considerably the purely electroweak
AWMM, this effect does not translate into an appreciable change of the
total AWMM (for which only  a 0.4\% of variation is found if $M_\Phi$
moves from $M_Z$ to $3\,M_Z$) because the electroweak contribution is less
than 1\% (for $M_\Phi=3M_Z$) of the total one.

Due to the fact that there is no enhancement of 
the electroweak contributions coming
from the presence of a heavy $top$-quark, all the electroweak diagrams 
(except those already mentioned with Higgs exchanged between two $b$'s) 
are of the same order, in particular the  $\gamma b \bar{b}$ diagram. Then, 
Eq. (\ref{qcd}) leads to the conclusion that the 
QCD contribution is  two orders of magnitude bigger than 
the electroweak one. 
In fact, the next to leading order contribution in perturbative QCD would be 
probably comparable to the
computed leading order in the electroweak sector.

For the $c$-quark, all the previous discussion holds, and
one expects the electroweak contribution to be of the order
$\frac{\alpha}{4\pi}\left( \frac{m_c}{M_Z}\right)^2$. That is
\beq
a^w_c(M_Z^2)\approx {\cal{O}}
\left(\left(\frac{m_c}{m_b}\right)^2\, a^w_b(M_Z^2)\right)\approx {\cal{O}}
\left(10^{-7}\right)
\eeq

The one-loop QCD contribution will also be  dominant, and its magnitude
can be easily computed from the analytic expression of Eq. (\ref{qcd}),
adapted to the $c$-quark. For $m_c=1.6$ GeV, we get the value:
\beq
a_c(M_Z^2)\approx a_c^{QCD}(M_Z^2)=\frac{\alpha_s}{\alpha}\frac{4}{3Q_c^2}
\left[a_c^w\right]^{\gamma c c}=(-2.80+1.09i)\times 10^{-5}
\eeq

\section{Conclusions}
We have calculated the electroweak contributions to the anomalous weak
magnetic moment of
the $b$-quark, within the Standard
Model, and found that it is of the order $10^{-6}$. One 
loop QCD contributions
to the AWMM are dominant and increase its value to 
$10^{-4}$. The result tells us that in the magnetic moment
form factor: 1) the contributions from new physics to the electroweak sector 
are hidden by the dominant strong
interaction contribution,  2) the $Zb\bar{b}$ width is rather insensitive to
electroweak contributions in the AWMM sector, and 3) contrary to what
happens for the charge radius form factor, non-decoupling effects do not take
place in the AWMM. The  value of the AWMM for 
the $c$-quark is also computed (up to first order in QCD)
and it is, as expected, smaller than that for the $b$-quark  by a factor 
$(m_c/m_b)^2 \times v_c/v_b$.

\section*{Acknowledgments}
We would like to thank Denis Comelli, Germ\'an Rodrigo and Arcadi 
Santamaria for clarifying discussions. 
This work has been supported in part by CICYT, under
Grant AEN 96-1718, by I.V.E.I., and by CONICYT under Grant 1039/94. 
G.A.G.S. thanks the Spanish Ministerio de Educaci\'on y 
Ciencia for a postdoctoral grant at the University of Valencia at the 
beginning of this work and  the hospitality received 
at the I.C.T.P. where  this work was completed.

\renewcommand{\thesection}{\Alph{section}}
\renewcommand{\theequation}{\thesection.\arabic{equation}}
\addtocounter{section}{-2}
\addtocounter{equation}{-22}

\end{document}